\documentclass[12pt]{article}
\usepackage{epsfig}

\usepackage{amssymb}
\usepackage{amsmath}
\usepackage{amsfonts}

\usepackage{amsthm}

\theoremstyle{theorem}
\newtheorem{lem}{Lemma}

\newtheorem{conj}{Conjecture}

\theoremstyle{definition}

\newtheorem{pozn}{Remark}

\def\bp{\begin{proof}}
\def\ep{\end{proof}}

\def\be{\begin{equation}}
\def\ee{\end{equation}}
\def\ba{\begin{array}{c}}
\def\ea{\end{array}}
\def\p{\partial}
\def\ben{\[}
\def\een{\]}

\newcommand{\bea}{\begin{eqnarray}}
\newcommand{\eea}{\end{eqnarray}}

\newcommand{\bbr}{\br\!\br}
\newcommand{\kkt}{\kt\!\kt}
\newcommand{\pbr}{\prec\!}
\newcommand{\pkt}{\!\succ\,\,}
\newcommand{\kt}{\rangle}
\newcommand{\br}{\langle}
\begin{document}

\titlepage

\vspace{.35cm}

 \begin{center}{\Large \bf

An exactly solvable quantum-lattice model with a tunable degree of
nonlocality

  }\end{center}

\vspace{10mm}

 \begin{center}

 {\bf Miloslav Znojil}

 \vspace{3mm}
Nuclear Physics Institute ASCR,

250 68 \v{R}e\v{z}, Czech Republic

{e-mail: znojil@ujf.cas.cz}

\vspace{3mm}


\end{center}

\vspace{5mm}


\section*{Abstract}

A non-Hermitian $N-$site-lattice Hamiltonian $H$ with
Laguerre-polynomial right eigenvectors and real energies is made
self-adjoint in an {\em ad hoc} Hilbert space ${\cal H}^{(S)}$. The
necessary physical inner products are defined via alternative
$k-$parametric $(2k+1)-$diagonal metrics $\Theta=\Theta_{k}$
constructed, in closed form, at $k=0,k=1,k=2$ and $k=3$. The value
of $k$ is interpreted as a degree of non-locality of the model.

\newpage

\section{Introduction \label{I} }

\subsection{The Laguerre-polynomial quantum model \label{Ia} }

It is well known that the Laguerre polynomials\footnote{our present
denotation $L(n, a, x)$ and normalization $L(-1,a,x) = 0,\ L(0,a,x)
= 1$ are taken from the symbolic-manipulation langauge MAPLE
\cite{Maple}; up to a factor the same polynomials are denoted as
$L^a_n(x)$ in Ref.~\cite{Ryzhik} or as $L^{(a)}_n(x)$ in
Ref.~\cite{Stegun}} $L(n, a, z)$ may {\em formally} be arranged in
an infinite-dimensional Dirac-ket-like column vector
 \be
 |\psi^{(\infty)}\kt=\left[
 \begin {array}{c}
 L(0, a, z)\ (=1)\\
 L(1, a, z)\ (=a+1-z)\\
 L(2, a, z)\ \left [=\frac{1}{2}(a+1)(a+2)
 -(a+2)\,z+\frac{1}{2}\,z^2
 \right ]
 \\
 \vdots
 \end {array} \right]\,
 \label{jedna}
 \ee
which satisfies, line-by-line and at any complex $z\in \mathbb{C}$,
the infinite linear algebraic system of equations
 \be
 H^{(\infty)}(a)\, |\psi^{(\infty)}\kt= z\,
|\psi^{(\infty)}\kt\,
 \label{nawodel}
  \ee
which resembles the Schr\"{o}dinger equation of quantum theory and
where
 \be
   H^{(\infty)}(a)=\left[
\begin {array}{ccccc}
a+1&-1&0&0&\ldots\\
\noalign{\medskip}-a-1&a+3&-2&0&\ldots\\
\noalign{\medskip}0&-a-2&a+5&-3&\ddots\\
\noalign{\medskip}0&0&-a-3&a+7&\ddots\\
\noalign{\medskip}\vdots&\vdots&\ddots&\ddots&\ddots
\end {array} \right]\,.
 \label{wodel}
 \ee
One can immediately formulate at least three reasons against
$H^{(\infty)}(a)$ of Eq.~(\ref{wodel}) being  interpreted as a
quantum Hamiltonian:

\begin{itemize}

\item {(pragmatic reason)} the set of all complex $z\in \mathbb{C}$ does
not look like a good candidate for a phenomenologically meaningful
spectrum;

\item {(mathematical reason)}
in the current, ``friendly" Hilbert space $\ell^2$ (to be denoted by
symbol ${\cal H}^{(F)}$, cf. Appendix) the norm of the wave-function
candidate (\ref{jedna}) would be infinite;

\item {(physical reason)}
the candidate $H^{(\infty)}(a)$ for the Hamiltonian of the system is
manifestly non-Hermitian.

\end{itemize}

\subsection{The finite-dimensional Laguerre-polynomial quantum model \label{Ib} }

Let us replace the matrix-resembling array $H^{(\infty)}(a)$ by its
finite-dimensional, truncated version
 \be
H^{(N)}(a)=\left[ \begin {array}{ccccc}
a+1&-1&0&\ldots&0\\
\noalign{\medskip}-a-1&a+3&-2&\ddots&\vdots\\
\noalign{\medskip}0&-a-2&a+5&-3&0\\
\noalign{\medskip}\vdots&\ddots&\ddots&\ddots&-N+1\\
 0&\ldots&0\ \ \ &\!\!\!\!-a-N+1 &a+2N-1
\end {array} \right]\,.
\label{nasmodel}
 \ee
A return of the finite-dimensional descendant of Eq.~(\ref{nawodel})
 \be
 H^{(N)}(a)\, |\psi^{(N)}_n\kt= E^{(N)}_n(a)\,
|\psi^{(N)}_n\kt\,,\ \ \ \ n = 1, 2, \ldots, N-1
 \label{nawodelka}
  \ee
to the status of Schr\"{o}dinger equation finds a new support in a
few suddenly emerging encouragements:

\begin{itemize}

\item {(pragmatic encouragement)} after truncation, the set of the admissible
eigenvalues shrinks from the whole complex plane of $z$ to the
discrete and non-degenerate $N-$plet of strictly real zeros of the
elementary secular equation
 \be
 L(N, a, z)=0\,,\ \ \ z = E_n\,, \ \ \ n = 0, 1, \ldots, N-1\,;
 \label{secular}
 \ee

\item {(mathematical encouragement)}
after truncation (and in both the Hilbert spaces ${\cal H}^{(F,S)}$
defined in Appendix), the finite array
 \be
 |\psi_n^{(N)}(a)\kt=\left[
 \begin {array}{c}
 L(0, a, E_n)\\
 L(1, a, E_n)\\
 \vdots\\
 L(N-1, a, E_n)
 \end {array} \right]\,
 \label{eigvec}
 \ee
may very easily be normalized, say, to one.

\end{itemize}

 \noindent
An obstacle which seems to survive is the third, non-Hermiticity
point and counterargument. In what follows we intend to show that
this argument is misleading and that in a way summarized in Appendix
the Hermiticity of Hamiltonian $H^{(N)}(a)$ may be achieved and
guaranteed in a large number of alternative physical Hilbert spaces
${\cal H}^{(S)}$ where the inner products are defined in terms of
the so called metric operator $\Theta$. We shall also show that for
our particular model these spaces (i.e., the metric-operator
matrices) may be constructed by purely non-numerical means.

\begin{table}[h]
\caption{A sample of the dimension- and parameter-dependence of the
energy spectra $\{E_n(a)\}$ of Hamiltonian (\ref{nasmodel}).}
\label{pexp4}

\vspace{2mm}

\centering
\begin{tabular}{||c|c||c|c|c|c|c||}
\hline \hline
 \multicolumn{2}{||c||}{\rm parameters}& \multicolumn{5}{c||}{\rm energies}\\
 \hline
  $N$& $a$ & $E_0(a)$&$E_1(a)$ & \ldots& $E_{N-2}(a)$&$E_{N-1}(a)$\\
 \hline \hline
 6&1.0&
 0.5276681217& 1.796299810&\ldots&
 11.23461043&
 17.64596355\\
 &2.0&
 0.8899410156& 2.433144232&\ldots&
 12.60041387&
 19.26204255\\
 &3.0&
 1.296419203& 3.093998381&\ldots&
 13.94134537&
 20.83985455\\
 \hline
 9&1.0&
  0.3681784529& 1.243357962&\ldots&
       20.38218199& 28.11834338
 \\
 &2.0&
 0.6318537723& 1.712163195& \ldots&
 21.90120660& 29.82533613
 \\
 &3.0&
 0.9343511232& 2.208578822& \ldots&
 23.39499254& 31.50012806
 \\
 \hline \hline
\end{tabular}
\end{table}

\subsection{A phenomenological addendum}

Our present solvable toy-model (\ref{nasmodel}) is just a very
special case of the broad class of the $N-$dimensional
tridiagonal-matrix Hamiltonians of the form
 \be
 \hat{H}^{(N)}=
  \left[ \begin {array}{cccccc}
   a_1&c_1&0&0&\ldots&0
  \\
  b_2
 &a_2&c_2&0&\ldots&0
 \\0&b_3&a_3&c_3&\ddots&\vdots
 \\0&0
 &\ddots&\ddots&\ddots&0
 \\\vdots&\ddots&\ddots&b_{N-1}&a_{N-1}&c_{N-1}
 \\{}0&\ldots&0&0&b_{N}&a_{N}\\
 \end {array} \right]\,
 \label{kitiel}
 \ee
which describe the one-dimensional single-particle $N-$site
quantum-lattice dynamics reduced to the mere nearest-neighbor
interaction. The kinematical aspect of virtually all of these models
$\hat{H}^{(N)}$ is, typically, characterized by the one-to-one
correspondence between the matrix index $s=1,2,\ldots,N$ and a
discrete coordinate $q_s$ of the ``site". In a more dimensional case
these sites may be numbered by an {\it ad hoc} multiindex, $s \to
\vec{s}$, but just the simplest one-dimensional case is to be
considered in what follows.

At the very start of our present application of the theoretical
innovation of the formalism using $\Theta \neq I$ in combination
with Hamiltonian~(\ref{kitiel}) and/or its special case
(\ref{nasmodel}) let us emphasize that such an effort may even be
supported by a purely pragmatic motivation. The numerical sample of
zeros $z = E_n^{(N)}(a)$ of Eq.~(\ref{secular}) as displayed in
Table~\ref{pexp4} persuades us, for example, that each of the
$N-$plets of the present toy-model bound-state energies really looks
very much like a realistic phenomenological spectrum.

Besides the similar intuitive argument it is equally important to
keep in mind that the necessary modifications and adaptations or
deformations of the spectrum are controlled not only by the freely
variable coupling {\it alias} self-coupling {\it alias}
asymmetry-strength $a>0$ but also by the very choice of the {\em ad
hoc} model-space dimension $N=1,2,\ldots$.

This means that even on the level of practical applicability, our
unusual model (\ref{nasmodel}) might find its place among the other
standard fitting tools for the description of experimental quantum
spectra, especially when they are not explained by the conventional
interactions yielding vibrations (like harmonic oscillator) or
rotations (mainly in more spatial dimensions). Thus, although the
core of our present message should be predominantly methodical and
mathematical, even its purely pragmatic aspects might prove relevant
and/or useful.


\section{The concept of position in quantum lattice\label{II}}

In the vast majority of phenomenological quantum-lattice models the
Hamiltonian is assumed Hermitian in the most common $N-$dimensional
vector space, i.e., we have $\hat{H}^{(N)}=\left(\hat{H}^{(N)}\right
)^\dagger$. In parallel, the coordinates $q_s$ are assumed real and
observable so that, in the Schr\"{o}dinger's ``mode of description"
\cite{Messiah}, the corresponding quantum operator of position is
most often represented by a diagonal and time-independent
$N-$dimensional matrix
 \be
 \hat{\mathfrak{q}}=
 \left(
 \begin{array}{cccc}
 q_1&&&\\
 &q_2&&\\
 &&\ddots&\\
 &&&q_{N}
 \ea
 \right )\,.
 \label{loc}
 \ee
Using the standard Dirac's notation the time-dependent
Schr\"{o}dinger equation then reads
 \be
 {\rm i}\p_t \,|\psi(t)\kt=\hat{H}^{(N)}\,|\psi(t)\kt
 \,
 \label{SEtD}
 \ee
and controls the evolution of the system prepared in a normalized
pure state at time $t_{prep}=0$. At the time of measurement $t>0$
one evaluates the wave function $\br q_s | \psi(t)\kt=\psi(t,s)$ and
determines the probability
  \be
 {\varrho}(t,s)=|\br q_s | \psi(t)\kt|^2=
  \psi^*(t,s)\,\psi(t,s)\,
  \label{SEtop}
  \ee
of finding the particle (or quasi-particle) at the $s-$th site
$q_s$.

\subsection{The loss of the measurability of position in the so
called ${\cal PT}-$symmetric Quantum Mechanics}

In the recent literature there emerged an interesting modification
of the whole paradigm (cf. its compact summary in  Appendix). It has
been developed, mainly, within the framework of the so called ${\cal
PT}-$symmetric quantum mechanics (cf., e.g., comprehensive reviews
\cite{Carl,ali} and also \cite{Geyer} or \cite{SIGMA}). In what
follows we intend to describe one of applications of this new
paradigm to our exactly solvable single-particle $N-$site
quantum-lattice model (\ref{nasmodel}) where the spectrum remains
real but where the Hamiltonian matrix itself appears {\em manifestly
non-Hermitian} in $\ell^2 \equiv {\cal H}^{(F)}$.

The technical essence of the new paradigm may be found summarized in
Appendix. {\it In nuce}, the key idea of the new formalism lies in
the weakening of the standard textbook Hermiticity $H  = H^\dagger$
of the Hamiltonian  and in its replacement by the requirement
 \be
 H^\dagger\,\Theta = \Theta\,H\,.
 \label{dieu}
 \ee
The use of this relation containing a nontrivial metric operator
$\Theta\neq I$ has already been shown useful, almost twenty years
ago, in nuclear physics \cite{Geyer}. Naturally, the same
requirement applies also to the other operators ${\cal O}$ of
observables where the rule ${\cal O}  = {\cal O}^\dagger$ must again
be replaced by its generalization
 \ben
  \Theta\,{\cal O}  = {\cal O}^\dagger\, \Theta\,.
  \een
In opposite direction, the ``lattice-position" matrix (\ref{loc})
loses its status of an observable quantity whenever the metric
ceases to be a diagonal matrix. This argument (nicely explained also
in Ref.~\cite{cubic})) further implies that also the quantity $
{\varrho}(t,s)$ of Eq.~(\ref{SEtop}) will lose its original
probabilistic interpretation in general.

\subsection{A return to the (smeared) positions. \label{physics}}

A partial resolution of the puzzle of the smearing of positions was
offered in Ref.~\cite{fund} where the concept of the position and
locality has {\em partially} been restored via the use of the ``next
to diagonal",  band-matrix, $(2k+1)-$diagonal matrices of the
metrics $ \Theta=\Theta_{k}=$
 \be
   =\left[ \begin {array}{ccccccc}
   \theta_{11}&\theta_{12}&\ldots&\theta_{1,k+1}&0&\ldots&0
  \\
  \theta_{21}
&\theta_{22}&\theta_{23}&\ldots&\theta_{2,k+2}&\ddots&\vdots
 \\
 \vdots&\ddots&\ddots&\ddots&&\ddots&0
 \\
 \theta_{k+1,1}&
 &\ddots&\ddots&\ddots&&\theta_{N-k,N}
 \\
 0&\ddots&
 &\ddots&\theta_{N-2,N-2}&\theta_{N-2,N-1}&\vdots
 \\
  \vdots&\ddots&\theta_{N-1,N-k-1}
  &\ldots&\theta_{N-1,N-2}&\theta_{N-1,N-1}&\theta_{N-1,N}
 \\ 0&\ldots&0&\theta_{N,N-k}&\ldots&\theta_{N,N-1}&\theta_{NN}
 \end {array} \right]\,.
 \label{kit}
 \ee
The integer $k = 0, 1, \ldots, N-1$ has been interpreted there as a
measure of the ``smearing". In this role, its value must be kept
perceivably smaller than the dimension $N$.

The physical meaning of the above non-diagonal metrics may be
clarified using the first nontrivial $k=1$ example in which we may
rewrite metric (\ref{kit}) as a superposition of a positive diagonal
matrix ${\cal D}^2$ with and upper-diagonal matrix $\alpha\,A$ and
its lower-diagonal conjugate $\alpha\,A^\dagger$,
 \be
 \Theta^{(N)}_1(a,\alpha)={\cal D}^2+\alpha\,\left (A+A^\dagger\right
 )\,.
 \label{tragm}
 \ee
Once we assume that the parameter $\alpha$ itself is sufficiently
small, we may recall formula (\ref{factorized}) of our Appendix and
deduce, say, the following approximate, non-Hermitian form of the
Dyson's map,
 \be
 \Omega =  \sqrt{\Theta^{(N)}_1(a,\alpha)}=
 {\cal D}+\frac{1}{2}\,\alpha\,{\cal D}^{-1}
 \left (
 A+A^\dagger
 \right ) +{\cal O}(\alpha^2)\,.
 \label{forfor}
  \ee
In this setting, it is still most natural to follow the notation of
our Appendix and to define the operator of position of our
(quasi)particle by the strictly diagonal matrix (\ref{loc}) acting
in the ``inaccessible" physical Hilbert space ${\cal H}^{(P)}$ where
the metric remains trivial, $\Theta^{(P)}=I$. Naturally, such a
position operator $ \hat{\mathfrak{q}}$ may be equally well
represented by the nondiagonal operator
 \ben
 \hat{Q}= \Omega^{-1}\hat{\mathfrak{q}}\Omega
 \een
acting either in the ``accessible", unitarily equivalent physical
Hilbert space ${\cal H}^{(S)}$ (where, in our notation,
$\hat{Q}=\hat{Q}^\ddagger$ is Hermitian) or in the unphysical and
unitarily non-equivalent auxiliary Hilbert space ${\cal H}^{(F)}$
(where the same operator remains non-Hermitian of course,
$\hat{Q}\neq \hat{Q}^\dagger$).

At this point our approximate formula (\ref{forfor}) for $\Omega$
enables us to transform $\hat{\mathfrak{q}}$ into its image
 \ben\
 \hat{Q} =
 \left [{\cal D}-\frac{1}{2}\,\alpha\,{\cal D}^{-1}
 \left (
 A+A^\dagger
 \right ) +{\cal O}(\alpha^2)
  \right ]
  \hat{\mathfrak{q}}
  \left [
  {\cal D}+\frac{1}{2}\,\alpha\,{\cal D}^{-1}
 \left (
 A+A^\dagger
 \right ) +{\cal O}(\alpha^2)
   \right ]=
 \een
 \ben
 ={\cal D}\hat{\mathfrak{q}}{\cal D}+
 \frac{1}{2}\,\alpha
 \left [{\cal D}
 \hat{\mathfrak{q}}{\cal D}^{-1}
 \left (
 A+A^\dagger
 \right )
 -{\cal D}^{-1}
 \left (
 A+A^\dagger
 \right )\hat{\mathfrak{q}}{\cal D}
 \right ]+{\cal O}(\alpha^2)
 \een
which is, up to second-order corrections, a diagonally dominated
tridiagonal matrix. The construction remains also feasible at $k>1$,
provided only that we make use of a partitioning and replace the
tridiagonality of $\Theta^{(N)}_1$ by the block-tridiagonality of
$\Theta^{(N)}_k$.

We may conclude that in our {\em ad hoc} physical Hilbert space
${\cal H}^{(S)}$ assigned to the cryptohermitian Hamiltonian
$H^{(N)}(a)$ the original basis $|q_s\kt$ [such that $\left
(|q_s\kt\right )_{s'} \sim \delta_{ss'}$] lost its connection with
the observable position. The role of the position-eigenstates is
taken over by the kets $|\chi_{q_s}\kt$ representing our
(quasi)particle(s) localized at a site $s$ (with spatial coordinate
$q_s$). They must be constructed as  eigevectors of our
cryptohermitian operator of position,
 \be
 \hat{Q}\,|\chi_{q_s}\kt= q_s \,|\chi_{q_s}\kt\,,\
  \ \ \ \ s = 1, 2, \ldots,N\,.
 \ee
This is a numerical problem, possibly simplified at small $\alpha$
when the operator of position predominantly couples the
nearest-neighbor basis states.

In the language of measurements  the probability of finding the
particle (or quasi-particle) in question at the $s-$th site of the
lattice may still be determined by formula~(\ref{SEtop}), after its
appropriate amendment of course. The form of this formula itself may
even stay unchanged, provided only that we replace the old,
Hermiticity-assuming definition of the wave function $\psi(t,s):=\br
q_s | \psi(t)\kt$ (meaning the ``probability-density quantity") by
its new, cryptohermiticity-assuming update
  \be
  \psi_{(\Theta_k)}(t,s)=\br \chi_{q_s} | \Theta_k |\psi(t)\kt=
  \sum_{s'=s-k}^{s+k}\,\br \chi_{q_s} | \left (\Theta_k \right )_{ss'}
   \left (|\psi(t)\kt
  \right )_{s'}
  \,.
  \ee
With the time-independent parameter $a \neq a(t)$ this definition
acquires, in the light of Eqs.~(\ref{secular}) and (\ref{eigvec}),
the final compact form
  \be
  \psi_{(\Theta_k)}(t,s)=
  \sum_{s'=s-k}^{s+k}\,\br q_s | \left (\Theta_k \right )_{ss'}
   \sum_{n=1}^N\,e^{iE_nt}\,L(s',a,E_n)  \,.
   \label{fina}
  \ee
This conclusion establishes the direct link between our
cryptohermitian Hamiltonian (\ref{nasmodel}) and the measurement od
the position. Formula (\ref{fina}) indicates that the different
choices of the physical metric will lead to the nonequivalent
predictions of the results of the experiment. {\em Vice versa}, the
experiments may be used, in principle, as a valid source of
information fitting the parameters in phenomenological metrics
\cite{Geyer}.

\section{A few mathematical prerequisites}

On the level of the textbook quantum theory the main obstacle of
accepting  the input Hamiltonian matrix $H^{(N)} \neq \left
(H^{(N)}\right )^\dagger$ (with real spectrum)  lies in the
technical nontriviality of making it compatible with the standard
postulates of Quantum Mechanics. In Appendix  we explained that this
acceptance is based on the ``hidden" Hermiticity
 \be
 H^{(N)}(a) = \left (H^{(N)}(a)\right )^\ddagger
 \label{dieube}
 \ee
(a.k.a. cryptohermiticity \cite{SIGMA}). This property is defined in
the less common but still entirely standard Hilbert space ${\cal
H}^{(S)}$. It may almost always be reread as the Dieudonn\'e's
\cite{Dieudonne} constraint (\ref{dieu}) imposed in the auxiliary,
friendlier Hilbert space ${\cal H}^{(F)}$.

For our model (\ref{nasmodel}) the major technical difficulties
related to the necessary construction of the right-hand-side
matrices in Eq.~(\ref{dieube}) (i.e., of the eligible metrics) will
be tractable due to the important observation that  the metrics may
{\em systematically} be constructed as special, $(2k+1)-$diagonal
matrices (\ref{kit}) at {\em any} $k \in \{0,1,\ldots, N-1\}$.

\subsection{The exceptional local model at $k=0$  \label{nulty}}

We shall assume that the parameter $a$ and the dimension $N$ are
both variable and, in principle, arbitrary while the number $k$ of
the nonvanishing diagonals in Eq.~(\ref{kit}) will remain, for
technical as well as interpretation reasons, fixed. Naturally, the
constructive analysis of the model should start from the trivial
$k=0$. In this case it is well known that within the formalism
described in Appendix, {\em any} tridiagonal complex matrix
(\ref{kitiel}) may be made compatible with the Dieudonn\'{e}'s
Hermitization equation $\hat{H}^\dagger \Theta = \hat{H}\,\Theta$
via the diagonal ansatz
 \be
  \Theta_{0}(a)= \left[ \begin {array}{ccccc}
   \theta_{11}&&&&
  \\
&\theta_{22}&&&
 \\
 &&\ddots&&
 \\
  &&&\theta_{N-1,N-1}&
 \\ &&&&\theta_{NN}
 \end {array} \right]\,
 \label{kit0}
 \ee
for the metric with the positive matrix elements, $\theta_{jj}>0$,
$j=1,2,\ldots,N$. This observation leads to the following easy
consequence.

\begin{lem}
Hamiltonians $H^{(N)}(a)$ of Eq.~(\ref{nasmodel}) may be assigned
the diagonal metrics $\Theta_{0}(a)$ of Eq.~(\ref{kit0}) with
normalization $\theta _{11}=1$ and with the elementary matrix
elements
 \be
  \theta _{nn}
                                  =
{\frac {(n-1)!}{ \left( a+n-1 \right)  \left( a+n-2 \right) \ldots
\left( a+2 \right)
 \left( a+1 \right) }}
 \label{k0}
 \ee
at $n = 2,3,\ldots,N$. \label{lemma1}
\end{lem}
\bp In the Dieudonn\'{e}'s equation $\hat{H}^\dagger \Theta =
\hat{H}\,\Theta$ it is easy to evaluate the left-hand-side product
of matrices for {\em any} tridiagonal  input (\ref{kitiel}),
 \be
 \left (\hat{H}^{(N)} \right )^\dagger\,\Theta=
  \left[ \begin {array}{cccccc}
   a_1^*\theta_{11}&b_2^*\theta_{22}&0&\ldots&0&0
  \\
  c_1^*\theta_{11}
  &a_2^*\theta_{22}&b_3^*\theta_{33}&0&\ldots&0
 \\0&c_2^*\theta_{22}&a_3^*\theta_{33}&b_4^*\theta_{44}&\ddots&\vdots
 \\\vdots&\ddots
 &\ddots&\ddots&\ddots&0
 \\{}0&\ldots&0&c_{N-2}^*\theta_{N-2,N-2}&a_{N-1}^*\theta_{N-1,N-1}&b_{N}^*\theta_{NN}
 \\{}0&\ldots&0&0&c_{N-1}^*\theta_{N-1,N-1}&a_{N}^*\theta_{NN}\\
 \end {array} \right]\,
 \label{kitiellev}
 \ee
and to compare it with the  right-hand-side tridiagonal matrix
 \be
 \Theta\,\hat{H}^{(N)}=
  \left[ \begin {array}{cccccc}
   \theta_{11}a_1&\theta_{11}c_1&0&0&\ldots&0
  \\
 \theta_{22} b_2
&\theta_{22}a_2&\theta_{22}c_2&0&\ldots&0
 \\0&\theta_{33}b_3&\theta_{33}a_3&\theta_{33}c_3&\ddots&\vdots
 \\0&\ddots
 &\ddots&\ddots&\ddots&0
 \\{}0&\ldots&0&\theta_{N-1,N-1}b_{N-1}&\theta_{N-1,N-1}a_{N-1}&\theta_{N-1,N-1}c_{N-1}
 \\{}0&\ldots&0&0&\theta_{NN}b_{N}&\theta_{NN}a_{N}\\
 \end {array} \right]\,.
 \label{kitielle}
 \ee
The net result of this comparison is that the diagonal elements of
the Hamiltonian must be real and that the sequence of relations
 \be
 \theta_{n+1,n+1}b_{n+1}=\theta_{nn}c_{n}^*\,,\ \ \ \ n = 1, 2,
 \ldots ,N-1\,
 \label{recurv}
 \ee
must be satisfied. For our present model (\ref{nasmodel}) this
observation immediately leads to the proof of the Lemma.
 \ep
\begin{pozn}

The simplicity of the above result made it useful in tests of a
straightforward symbolic-manipulation algorithm which we intended to
use for the computer-assisted solution of the Dieudonn\'{e}'s
equation $\hat{H}^\dagger \Theta = \hat{H}\,\Theta$ using the
various less elementary forms of the ansatz for $(2k+1)-$diagonal
metric $\Theta=\Theta_k$. Identifying the present model with the
$k=0$ special case the algorithm produced the sequence of formulae
   \ben \theta_{22}
                                  =
{\frac {\theta_{11}}{a+1}}
                            \,,\ \ \ \  \theta_{33}
                                  =
{\frac {2\,\theta_{11}}{ \left( a+2 \right)  \left( a+1 \right)
}}\,, \ \ \ldots
                             \een
and confirmed the validity of the above closed-form rigorous result
(\ref{k0}) and, {\it ipso facto}, also the reliability of the
algorithm itself. \label{poznamka1}
\end{pozn}

\subsection{The left eigenvectors of $H^{(N)}(a)$ at $k > 1$}

As long as we have to work with the phenomenological model of
quantum dynamics where $\hat{H}^{(N)}(a) \neq
\left(\hat{H}^{(N)}(a)\right )^\dagger$ at all $N$ and $a$, it is
insufficient to know just the {\it a priori} specified solutions
(\ref{eigvec}) of the current time-independent eigenvalue problem
(\ref{nawodelka}). For multiple purposes it is also necessary to
know the left eigenvectors $\br \xi_n|$ of our Hamiltonian. Most
often, we rather construct their duals  $| \xi_n\kt$ defined as the
right, usual eigenvectors of the conjugate, unusual Hamiltonian
$\left(\hat{H}^{(N)}(a)\right )^\dagger$,
 \be
 \left (H^{(N)}(a)\right )^\dagger\,
 |\xi^{(N)}_n\kt= E^{(N)}_n(a)\,
 |\xi^{(N)}_n\kt\,,\ \ \ \ n = 1, 2, \ldots, N-1\,.
 \label{nawodelkabe}
  \ee
As long as we already know the energies, this is a simpler task. A
rather subtle problem only emerges when we recollect that a suitable
set of the left eigenbras $\bbr \psi_n^{(N)}|$ of matrix
$H^{(N)}(a)$ has already been defined in Appendix (cf.
Eq.~(\ref{newcon})). This implies the necessity of the following
proportionality relation
 \be
 |\psi_n^{(N)}\kkt := \Theta\,|\psi_n^{(N)}\kt
 =|\xi^{(N)}_n\kt\,\kappa_n^{(N)}(a)\,
 \label{kappas}
 \ee
between the two alternative sets of the eigenvectors where the
left-hand-side-state normalization is fixed while the
right-hand-side-state normalization still remains open in
Eq.~(\ref{nawodelkabe}).

One of the consequences of the arbitrariness of the $n-$dependent
complex constants $\kappa_n^{(N)}(a)$ is particularly serious since
the solutions $|\xi^{(N)}_n\kt$ of Eq.~(\ref{nawodelkabe}) are often
used in the spectral-expansion definition of the metric
\cite{pseudo,Ali},
 \be
 \Theta= \sum_{n=0}^{N-1}\,|\psi_n^{(N)}\kkt \,\bbr\psi_n^{(N)}|=
 \sum_{n=0}^{N-1}\,|\xi_n^{(N)}\kt \,\left |\kappa_n^{(N)}(a) \right |^2
 \,\br\xi_n^{(N)}|\,.
 \label{hu}
 \ee
This implies that up to an inessential overall multiplication factor
(say, $\left |\kappa_0^{(N)}(a) \right |^2>0$), the metric
(\ref{hu}) contains, in general, an $(N-1)-$plet of free parameters
$\left |\kappa_n^{(N)}(a) \right |^2>0$, $n=1,2,\ldots,N-1$. As a
consequence, {\em every} given Hamiltonian $H^{(N)}(a)$ may be
assigned an $(N-1)-$parametric multiplet of alternative metrics
leading, generically, to independent physical predictions of the
model.

For this reason it is important to have, always, a sensible {\em
physics-based} recipe for an efficient suppression of the latter
ambiguity. In field theory one often requires the observability of
the so called ``charge" ${\cal C}$ for this purpose \cite{Carl}. For
quantum lattices using the nearest-neighbor interaction dynamics we
proposed, very recently \cite{fund}, another principle of
suppression of the ambiguity of the metrics which is also being used
here. This recipe is based on the requirement of the minimal (or at
least ``tunable", controllable) non-locality of the model in
question. In the latter setting, unfortunately, the {\em direct} use
of spectral formula (\ref{hu}) does not work at all. At the same
time, the use of the metric-multiplication definition (\ref{kappas})
of the left eigenvectors $|\psi_n^{(N)}\kkt$ may be recommended as
facilitated by the $(2k+1)-$diagonal band-matrix structure of the
metric, especially at the not too large $k$.

\section{The
method of construction of the multi-diagonal metrics}

\subsection{The sparse-matrix-expansion ansatz.}

Our preliminary computer-assisted experiments with solving
Dieudonn\'{e}'s Eq.~(\ref{dieu}) with Hamiltonian (\ref{nasmodel})
revealed that its $(2k+1)-$diagonal matrix solutions (\ref{kit})
might be sought in a specific linear-superposition form
 \be
 \Theta=\Theta^{(N)}_{k}(a)=
 \Theta^{(N)}_{(k,\alpha1,\alpha_2,\ldots,\alpha_k)}(a)=
 \Theta^{(N)}_0(a)+\sum_{j=1}^k\,\alpha_j\,{\cal P}_j^{(N)}(a)\,
 \label{every}
 \ee
with $k \in \{1,2,\ldots,N-1\}$ and with  the diagonal metric of
paragraph \ref{nulty} accompanied by a $k-$plet of $(2j+1)-$diagonal
real and symmetric matrices ${\cal P}_j^{(N)}(a)$. In fact, just the
diagonal matrix $\Theta^{(N)}_0(a)$ is a true, positive-definite
metric. Every other matrix ${\cal P}_j^{(N)}(a)$ is only expected
{\em individually} compatible with the Dieudonn\'{e}'s constraint,
 \be
  \left (H^{(N)}(a)\right )^\dagger\,{\cal P}_j^{(N)}(a) ={\cal
 P}_j^{(N)}(a)\,H^{(N)}(a)\,,\ \ \ \ \ j = 1, 2, \ldots, k\,.
 \label{dieucet}
 \ee
In this overall framework the core of the exhaustive construction of
the expansion (\ref{every}) at a given non-locality $k$ lies,
obviously, {\em in} the constructions of all of the pseudometrics
(\ref{dieucet}) with $j \leq k$ {\em and in} having the real
expansion parameters $\alpha_j$ sufficiently small to keep the
necessary positive definiteness of the diagonally-dominated sum
(\ref{every}) unbroken.

\subsection{The extrapolation method of solving Eq.~(\ref{dieucet}).}

In what follows the systematic and explicit construction of
pseudometrics ${\cal P}_j^{(N)}(a)$ will be performed for all $j
\leq 3$. This construction will proceed in three steps.

In a preparatory step we select an integer value of subscript $j\geq
1$, insert a general $(2j+1)-$diagonal-matrix ansatz for ${\cal
P}_j^{(N)}(a)$ in Eq.~(\ref{dieucet}) and, using a computer-assisted
trial-and-error strategy, fine-tune an ansatz for ${\cal
P}_j^{(N)}(a)$ in such a way that a maximum of its matrix elements
not lying on its outer diagonals is being set equal to zero.

In the second step of the algorithm we select a few $N\leq N_0$ and,
via the computer-assisted closed-form solution of
Eq.~(\ref{dieucet}) we construct {\em all} of the corresponding
matrix elements $\theta_{mm'}=\theta_{mm'}^{(N,j)}(a)$ of ${\cal
P}_j^{(N)}(a)$. In our concrete model this computer-assisted
``experiment" revealed, after certain symbolic-manipulation
factorizations and simplifications of the originally quite tedious
results, the prevailing $N-$independence and sufficiently
transparent and nicely factorizable $a-$dependence of the individual
matrix elements of our separate pseudometric matrices ${\cal
P}_j^{(N)}(a)$, at the first few smallest $j$ at least.

In the third step the resulting sample of elements must be perceived
and reanalyzed as a set which is generated by the comparatively
elementary linear algebraic set of equations (\ref{dieucet}). These
equations always possess, at any fixed $N$ and $j$, an implicitly
recurrent structure. Such a reinterpretation of the algebra opens
the possibility of skipping the intermediate step of the {\em
explicit} determination of the individual recurrences as sampled, at
$j=0$, by Eq.~(\ref{recurv}) above. Especially at the higher values
of $j$ such an intermediate step did prove prohibitively complicated
and, at the same time, redundant, especially due to the elementary
nature of our Hamiltonian in question.

Thus, in the third step we replace the tedious multidimensional
recurrent generation of the closed algebraic formulae for the
unknown multiplets of elements $\theta_{mm'}$ by the extrapolation
technique. In our model the latter recipe proved more efficient than
the direct use of recurrences even at the lowest values of $j$.
Naturally, having the closed form of extrapolation at our disposal
at last, the final rigorous proof of its general compatibility with
Eq.~(\ref{dieucet}) by direct insertion remains straightforward.

\section{The model with minimal smearing ($k=1$).\label{IV}}


In the general family of $\alpha-$dependent tridiagonal metrics
 \be
 \Theta^{(N)}_1(a,\alpha)=\Theta^{(N)}_0(a)+\alpha\,{\cal P}^{(N)}_1(a)\,
 \label{tridiagm}
 \ee
``numbered" by a real and not too large variable $\alpha$ we may
assume, without any loss of generality, that the matrix ${\cal
P}^{(N)}_1(a)$ is real and symmetric and that without loss of
generality its element $\theta_{11}$ may be chosen as vanishing,
 \be
 {\cal P}^{(N)}_1(a) = \left[ \begin {array}{ccccccc}
   0&\theta_{12}&&&&&
  \\
  \theta_{12}
&\theta_{22}&\theta_{23}&&&&
 \\
 & 
  \theta_{23}
&\theta_{33}&\theta_{34}&&&
 \\
 &
 &\ddots&\ddots&\ddots&&
 \\
 &&
 &\theta_{N-3,N-2}&\theta_{N-2,N-2}&\theta_{N-2,N-1}&
 \\
  &&
  &&\theta_{N-2,N-1}&\theta_{N-1,N-1}&\theta_{N-1,N}
 \\ &&&&&\theta_{N-1,N}&\theta_{NN}
 \end {array} \right]\,.
 \label{kit1}
 \ee
One of the most unexpected observations made during the brute-force
construction of this pseudometric appeared to be the
cutoff-independence of the result and, in particular, of the ``last"
element $\theta_{NN}\neq 0$. Thus the ``initial" list of the
elements
  \ben \theta_{22}
                                  =
-{\frac {2}{a+1}}
                             \,,\ \ \  \theta_{33}
                                  =
-{\frac {8}{ \left( a+2 \right)  \left( a+1 \right) }}
                             \,,
                             \een \ben \theta_{44}
                                  =
-{\frac {36}{ \left( a+3 \right)  \left( a+2 \right)  \left( a+1
 \right) }}\,,\ \ \  \theta_{12}
                                  =
1\,,
                             \een
  \ben \theta_{23}
                                  =
{\frac {2}{a+1}}
                             \,,\ \ \  \theta_{34}
                                  =
{\frac {6}{ \left( a+2 \right)  \left( a+1 \right) }}
                             \een
obtained at $N\leq N_0=4$ proved sufficient for the extrapolation
and for the final formulation of the general result.

\begin{lem}
Hamiltonians $H^{(N)}(a)$ of Eq.~(\ref{nasmodel}) may be assigned
the tridiagonal family of metrics (\ref{tridiagm}). The elementary
matrix elements of pseudometric (\ref{kit1}) are given by the
cutoff-insensitive formulae
 \ben
  \theta_{nn}
                                  =
{\frac {(n-1)!}{ \left( a+n-1 \right)  \left( a+n-2 \right) \ldots
\left( a+2 \right)
 \left( a+1 \right) }}
 \label{k0b}
 \een
at $n = 2,3,\ldots,N$, and
 \ben   \theta_{nn+1}
                                  =
{\frac {n!}{ \left( a+n-1 \right)  \left( a+n-2 \right) \ldots
\left( a+2 \right)
 \left( a+1 \right) }}
 \een
at $n = 1,2,\ldots,N-1$. \label{lemma2}
\end{lem}
\bp In a parallel to the constructive proof of Lemma \ref{lemma1}
the tridiagonal-metric-generated recurrent relations remain
sufficiently transparent to admit the explicit identification and
the straightforward proof of validity of their closed solutions by
mathematical induction not only for our model (\ref{nasmodel}) but
for any tridiagonal input Hamiltonian (\ref{kitiel}). The details
are left to the readers.
 \ep

\subsection{The verification of extrapolation hypotheses}

The computation technique which we use here may be characterized as
an interactive algorithm based on a systematic computer-assisted
verifications of the series of amended extrapolation hypotheses. It
will remain applicable, {\em mutatis mutandis}, at any $k > 1$,
throughout our forthcoming constructions.

The key technical problem will always lie in the determination of an
appropriate ansatz for pseudometrics. In practice, the only way of
finding the optimal ansatz seems to lie in a patient, brute-force
solution of Eq.~(\ref{dieu}) at a sequence of the smallest
dimensions $N$. That's what we will always have to do. Without the
help of MAPLE, such a task would be rather difficult.

In an illustrative verification of the result presented in Lemma
\ref{lemma2} we may compare its predictions, say, with their two
computer-generated counterparts
                           \ben \theta_{45}
                                  =
{\frac {24}{ \left( a+3 \right)  \left( a+2 \right)  \left( a+1
 \right) }}
                             \,,\ \ \  \theta_{55}
                                  =
-{\frac {192}{ \left( a+4 \right)  \left( a+3 \right)  \left( a+2
 \right)  \left( a+1 \right) }}\,.
                             \een
This comparison reconfirms the validity of Lemma \ref{lemma2}.
Moreover, these and similar ``redundant" formulae may offer an
insight and background for the estimates, say, of the numerical
magnitude of the eigenvalues of the metric at larger parameters
$\alpha$ and/or $a$. In the context of a different quantum-lattice
model a confirmation of feasibility of such a nonperturbative search
for the strong-coupling boundaries of the domain of positivity of
the metric was mediated, e.g., by Table I of Ref.~\cite{Gegenb}.

An additional merit of the interactive amendment of the tentatively
extrapolated formulae via the {\em a posteriori} computer-assisted
verification has been found in its speed. The initial tedious
algorithm of the lengthy direct construction gets easily amended in
the light of  the extrapolation so that the calculations at the
higher dimensions prove, paradoxically, quicker. Giving, in our
illustrative example, the sequence of  further elements
 \ben \theta_{56}
                                  =
{\frac {120}{ \left( a+4 \right)  \left( a+3 \right)  \left( a+2
 \right)  \left( a+1 \right) }}\,,
                             \een  \ben \theta_{66}
                                  =
-{\frac {1200}{ \left( a+5 \right)  \left( a+4 \right)  \left( a+3
 \right)  \left( a+2 \right)  \left( a+1 \right) }}
                             \een
(etc) which further confirm the reliability of the formulae and open
the way to the perceivably simplified proofs using direct
insertions.

\section{Pentadiagonal metrics ($k=2$) \label{V} }

In the general pentadiagonal metric of our model (\ref{nasmodel}),
 \be
 \Theta^{(N)}_2(a,\alpha,\beta)=\Theta^{(N)}_0(a)+\alpha\,{\cal P}^{(N)}_1(a)
 +\beta\,{\cal P}^{(N)}_2(a)\,\,
 \label{pentadiagm}
 \ee
the only unknown matrix ${\cal P}^{(N)}_2(a)$ must again be real and
symmetric, normalized, say, by
       the choice of $ \theta(1, 3)=1$ and finally, without loss of
       generality,
        admitting that the elements $
       \theta_{11}$ and $ \theta_{12}$ vanish,
 \be
 {\cal P}^{(N)}_2(a) = \left[ \begin {array}{ccccccc}
   0&0&\theta_{13}&&&&
  \\
  0
&\theta_{22}&\theta_{23}&\theta_{24}&&&
 \\
 \theta_{13}& 
  \theta_{23}
&\theta_{33}&\theta_{34}&\theta_{35}&&
 \\
 &\ddots
 &\ddots&\ddots&\ddots&\ddots&
 \\
 &&\theta_{N-4,N-2}
 &\theta_{N-3,N-2}&\theta_{N-2,N-2}&\theta_{N-2,N-1}&\theta_{N-2,N}
 \\
  &&
  &\theta_{N-3,N-1}&\theta_{N-2,N-1}&\theta_{N-1,N-1}&\theta_{N-1,N}
 \\ &&&&\theta_{N-2,N}&\theta_{N-1,N}&\hat{t}^{(N)}_{NN}
 \end {array} \right]\,.
 \label{kit2}
 \ee
The hat-superscripted matrix element $\hat{t}^{(N)}_{NN}$ is
exceptional and must be considered manifestly cutoff-dependent. This
has been revealed during the first-step calculations which gave the
initial list of the cutoff-insensitive elements
    \ben  \theta_{22}
                                  =
{\frac { a+2  }{a+1}}
                             \,,\ \ \   \theta_{33}
                                  =
{\frac {4\,\left( a+5 \right) }{ \left( a+2 \right)  \left( a+1
 \right) }}\,,\ \ \
                             \theta_{44}
                                  =
{\frac {18\, \left( a+8 \right) }{ \left( a+3 \right)  \left( a+2
 \right)  \left( a+1 \right) }}
                             \,, \een
 \ben   \theta_{55}
                                  =
{\frac {96\, \left( a+11 \right) }{ \left( a+4 \right)  \left( a+3
 \right)  \left( a+2 \right)  \left( a+1 \right) }} \neq
 \hat{t}^{(5)}_{55} =
{\frac {36\, \left( a+21 \right) }{ \left( a+4 \right)  \left( a+3
 \right)  \left( a+2 \right)  \left( a+1 \right) }}
 \,,
                             \een  \ben  \theta_{23}
                                  =-{\frac {4}{a+1}}
                             \,,\ \ \   \theta_{34}
                                  =
-{\frac {24}{ \left( a+2 \right)  \left( a+1 \right) }}
                             \,,\ \ \   \theta_{45}
                                  =
-{\frac {144}{ \left( a+3 \right)  \left( a+2 \right)  \left( a+1
 \right) }}\,,
                             \een
                             \ben  \theta_{13}
                                  =1
                             \,,\ \ \ \   \theta_{24}
                                  =
{\frac {3}{a+1}}
                            \,,\ \ \ \   \theta_{35}
                                  =
{\frac {12}{ \left( a+2 \right)  \left( a+1 \right) }}\,,
                             \een
leading, by extrapolation and by its subsequent tests at a few
$N>N_0=5$, to the following general result.

\begin{lem}
Hamiltonians $H^{(N)}(a)$ of Eq.~(\ref{nasmodel}) may be assigned
the pentadiagonal family of metrics (\ref{pentadiagm}). The
cutoff-insensitive matrix elements of pseudometric (\ref{kit2}) are
given by formula
  \ben  \theta_{nn}
                                  =
{\frac {(n-1)\,(n-1)! \left( a+3\,n-4 \right) }{ \left( a+n-1
\right) \left( a+n-2
 \right) \ldots    \left( a+2 \right)  \left( a+1 \right) }}
                             \label{k20}
 \een
at $n = 2,3,\ldots,N$, and by
 \ben
   \theta_{nn+1}
                                  =
-{\frac {2\,(n-1)\,n!}{  \left( a+n-1 \right) \left( a+n-2
 \right) \ldots    \left( a+2
 \right)
 \left( a+1 \right) }}
                              \label{k21}
 \een
at $n = 2,3,\ldots,N-1$, and by
 \ben
  \theta_{nn+2}
                                  =
{\frac {(n+1)!}{2\, \left( a+n-1 \right)  \left( a+n-2 \right)
\ldots \left( a+2 \right)  \left( a+1 \right) } }
 \label{k22}
 \een
at $n = 2,3,\ldots,N-2$. \label{lemma3}
\end{lem}
 \bp
We leave the proof by direct insertion to the readers.
 \ep
\begin{pozn}
The determination of the missing exceptional-element sequence
$\hat{t}^{(N)}_{NN}$ requires the use of a different approach,
outlined in section \ref{VIa} below.
\end{pozn}

\section{The metrics with seven  diagonals, $k=3$. \label{VI} }

In the seven-diagonal metrics
 \be
 \Theta^{(N)}_3(a,\alpha,\beta,\gamma)=\Theta^{(N)}_0(a)
 +\alpha\,{\cal P}^{(N)}_1(a)
 +\beta\,{\cal P}^{(N)}_2(a)+\gamma\,{\cal P}^{(N)}_3(a)\,\,
 \label{heptadiagm}
 \ee
the missing real and symmetric component ${\cal P}^{(N)}_3(a)$ may
be constructed along the same lines as above, starting from the
assumptions
        $
       \theta_{11}=\theta_{22}=0$, $ \theta(1, 2)= \theta(1, 3)=0$
       and
        $ \theta(1, 4)=1$ and from the
 heptadiagonal ansatz
  \be
 {\cal P}^{(N)}_3(a) = \left[ \begin {array}{ccccccc}
   0&0&0&\theta_{14}&&&
  \\
    0 &0&\theta_{23}&\theta_{24}&\ddots&&
 \\
 0&
  \theta_{23} &\theta_{33}&\theta_{34}&\ddots&\theta_{N-4,N-1}&
 \\
 \theta_{14}&
  \theta_{24} &\theta_{34}&\theta_{44}&\ddots&\theta_{N-3,N-1}&\theta_{N-3,N}
 \\
 &\theta_{25}
 &\ddots&\ddots&\ddots&\theta_{N-2,N-1}&\theta_{N-2,N}
 \\
  &&\ddots
  &\theta_{N-3,N-1}&\theta_{N-2,N-1}&\theta_{N-1,N-1}&\hat{t}_{N-1,N}
 \\ &&&\theta_{N-3,N}&\theta_{N-2,N}&\hat{t}_{N-1,N}&\hat{t}^{(N)}_{NN}
 \end {array} \right]\,.
 \label{kit3}
 \ee
We now encounter the two specific, hat-superscripted matrix elements
$\hat{t}^{(N)}_{NN}$ and $\hat{t}^{(N)}_{N-1,N}$ which must be
constructed in different manner (cf. section \ref{VIa} below).

For the purposes of extrapolation we evaluated, this time, the
following $N_0=6$ set of the cutoff-insensitive matrix elements,
 \ben \theta_{33}
                                  =
-8\,{\frac {a+3}{ \left( a+2 \right)  \left( a+1 \right) }}
                              \,,\ \ \  \theta_{44}
                                  =
-24\,{\frac {3\,a+14}{ \left( a+3 \right)  \left( a+2 \right) \left(
a+1 \right) }}\,,
                              \een
                              \ben \theta_{55}
                                  =
-192\,{\frac {3\,a+19}{ \left( a+4 \right)  \left( a+3 \right)
 \left( a+2 \right)  \left( a+1 \right) }}\,,
                              \een  \ben \theta_{66}
                                  =
-4800\,{\frac {a+8}{ \left( a+5 \right)  \left( a+4 \right)  \left(
a+ 3 \right)  \left( a+2 \right)  \left( a+1 \right) }}\,,
                              \een
                              \ben \theta_{23}
                                  =
{\frac {a+3}{a+1}}
                             \,,\ \ \  \theta_{34}
                                  =
6\,{\frac {a+8}{ \left( a+2 \right)  \left( a+1 \right) }}\,,
                              \een
                              \ben \theta_{45}
                                  =
36\,{\frac {a+13}{ \left( a+3 \right)  \left( a+2 \right)  \left(
a+1
 \right) }}\,,
                              \een \ben \theta_{56}
                                  =
240\,{\frac {a+18}{ \left( a+4 \right)  \left( a+3 \right)  \left(
a+2
 \right)  \left( a+1 \right) }}\,,
                              \een
  \ben \theta_{24}
                                  =
- \frac{6}{a+1}
                              \,,\ \ \  \theta_{35}
                                  =
-{\frac {48}{ \left( a+2 \right)  \left( a+1 \right) }}
                              \,,\ \ \  \theta_{46}
                                  =
-{\frac {360}{ \left( a+3 \right)  \left( a+2 \right)  \left( a+1
 \right) }}     \,,                         \een
  \ben \theta_{14}
                                  =
1
                              \,, \theta_{25}
                                  =
 \frac{4}{ a+1 }
                              \,,\ \ \  \theta_{36}
                                  =
{\frac {20}{ \left( a+2 \right)  \left( a+1 \right) }}\,
                              \een
leading, by extrapolation and its subsequent verification, to the
following general result.

\begin{lem}
Hamiltonians $H^{(N)}(a)$ of Eq.~(\ref{nasmodel}) may be assigned
the heptadiagonal family of metrics (\ref{heptadiagm}). The
cutoff-insensitive matrix elements of pseudometric (\ref{kit3}) are
given by formula
  \ben
          \theta_{nn}
                                  =
 -{\frac {2\,(n-1)\,(n-2)\,(n-1)!\, \left( a+5\,n-6 \right) }{3\, \left( a+n-1
\right) \left( a+n-2
 \right) \ldots    \left( a+2 \right)  \left( a+1 \right) }}
                    \label{k30}
 \een
at $n = 2,3,\ldots,N$, and by
 \ben
 \theta_{nn+1}
                                  =
{\frac {(n-1)\,n!\,(a+5\,n-7)}{2\,  \left( a+n-1 \right) \left(
a+n-2
 \right) \ldots    \left( a+2
 \right)
 \left( a+1 \right) }}
                              \label{k31}
 \een
at $n = 2,3,\ldots,N-1$, and by
 \ben
  \theta_{nn+2}
                                  = -
{\frac {(n-1)\,(n+1)!}{ \left( a+n-1 \right)  \left( a+n-2 \right)
\ldots \left( a+2 \right)  \left( a+1 \right) } }
 \label{k32}
 \een
at $n = 2,3,\ldots,N-2$, and by
 \ben
  \theta_{nn+3}
                                  =
{\frac {(n+2)!}{6\, \left( a+n-1 \right)  \left( a+n-2 \right)
 \ldots \left( a+2 \right)  \left( a+1 \right) } }
 \label{k33}
 \een
at $n = 2,3,\ldots,N-3$. \label{lemma4}
\end{lem}
 \bp
Again, we leave the proof by direct insertion to the readers.
 \ep

\section{The exceptional, cut-off-dependent matrix elements $\hat{t}^{(N)}$ \label{VIa}}

\subsection{ The construction of $\hat{t}^{(N)}_{NN}$ at $k=2$.}

The mathematical nature of the determination of the exceptional
elements in ansatz (\ref{kit2}) is special. Firstly, it is fairly
clumsy to think about their values as resulting from recurrences
since they vary with $N$ themselves. Secondly, the brute-force
collection of data needed for extrapolation with respect to $N$ is
much more time-consuming. At the same time, one can extrapolate
these data with respect to the integer variable $N$ almost as easily
as in the case of the other matrix elements.

In such a situation we decided not to search for the rigorous
proofs. Our readers will only be offered here the empirically
multiply reconfirmed (i.e., for us, credible enough) results of the
computer-assisted (though still rather time-consuming)
extrapolations.

\begin{conj}
At $j=2$ and at all $N = 3, 4, \ldots$ formula
 \ben        \tilde{t}^{(N)}_{NN}
                                  =
{\frac {(N-2)\,(N-1)!\,\left( a+5\,N-4 \right) }{2\, \left( a+N-1
\right) \left( a+N-2
 \right) \ldots   \left( a+2 \right)  \left( a+1 \right) }
 } \een
defines the ``last missing matrix element"  $\hat{t}^{(N)}_{NN}$ of
pseudometric (\ref{kit2}). \label{lemma5}
\end{conj}
\begin{pozn}
The validity of the latter formula (obtained by the extrapolation
from the data at $N\leq N_0=5$) has been made plausible, for us, by
the direct MAPLE-mediated evaluation of the subsequent values up to
  \ben  \tilde{t}^{(9)}_{99}
                                  =
{\frac {141120\, \left( a+41 \right) }{ \left( a+8 \right)  \left(
a+ 7 \right)  \left( a+6 \right)  \left( a+5 \right)  \left( a+4
\right)
 \left( a+3 \right)  \left( a+2 \right)  \left( a+1 \right) }}\,.
                             \een
\end{pozn}

\subsection{ The construction of $\hat{t}^{(N)}_{NN}$  at $k=3$.}

The computer-assisted solution of Eq.~(\ref{dieucet}) enabled us to
evaluate the following set of results of symbolic manipulations,
 \ben \hat{t}^{(4)}_{44}
                                  =
-16\,{\frac {a+7}{ \left( a+3 \right)  \left( a+2 \right)  \left(
a+1
 \right) }}
 \een

 \ben \hat{t}^{(5)}_{55}
                                  =
-16\,{\frac {11\,a+103}{ \left( a+4 \right)  \left( a+3 \right)
 \left( a+2 \right)  \left( a+1 \right) }}
 \een

                       \ben      \hat{t}^{(6)}_{66}
                                  =
-240\,{\frac {7\,a+82}{ \left( a+5 \right)  \left( a+4 \right)
 \left( a+3 \right)  \left( a+2 \right)  \left( a+1 \right) }}
 \een
                          \ben  \hat{t}^{(7)}_{77}
                                  =
 -960\,{\frac {17\,a+239}{ \left( a+6 \right)  \left( a+5 \right)
 \left( a+4 \right)  \left( a+3 \right)  \left( a+2 \right)  \left( a+
 1 \right) }}
 \een
which proved extrapolated as follows.

\begin{conj}
At $j=3$ and at all $N = 3, 4, \ldots$ formula
  \ben
                             \hat{t}^{(N)}_{NN}
                                  =-
{\frac {(N-3)\,(N-1)!\,[(3\,N-4)\,a+7\,N^2-16\,N+8]}{3 \left( a+N-1
\right) \left( a+N-2 \right) \ldots  \left( a+2 \right) \left( a+1
\right) }}
 \een
defines the diagonal missing matrix element  of pseudometric
(\ref{kit3}). \label{lemma6}
\end{conj}
\begin{pozn}
Again, for us, the plausibility of this conjecture has been enhanced
by the MAPLE-mediated evaluation of the subsequent values up to
 \ben
\hat{t}^{(9)}_{99}
                                  =
-80640\,{\frac {23\,a+431}{ \left( a+8 \right)  \left( a+7 \right)
 \left( a+6 \right)  \left( a+5 \right)  \left( a+4 \right)  \left( a+
3 \right)  \left( a+2 \right)  \left( a+1 \right) }}\,.
                              \een
A deeper meaning of the display of similar formulae may be seen not
only in the misprint-control in the general formulae but also in the
facilitated future identifications of the underlying recurrences if
any.  In addition, the  transparency of the formulae (and, in
particular, of their $a-$dependence) also underlines the formal
simplicity of the metric itself, especially when we compare its
matrix elements, say, with their complicated polynomial analogues as
obtained in Ref.~\cite{fund}.
\end{pozn}

\subsection{ The construction of $\hat{t}^{(N)}_{N-1N}$  at $k=3$.}

In a way paralleling the preceding section we started from the data
    \ben \hat{t}^{(4)}_{34}
                                  =
2\,{\frac {a+16}{ \left( a+2 \right)  \left( a+1 \right) }}
 \een
 \ben \hat{t}^{(5)}_{45}
                                  =
16\,{\frac {a+23}{ \left( a+3 \right)  \left( a+2 \right)  \left(
a+1
 \right) }}
 \een
  \ben        \hat{t}^{(6)}_{56}
                                  =
120\,{\frac {a+30}{ \left( a+4 \right)  \left( a+3 \right)  \left(
a+2
 \right)  \left( a+1 \right) }}
 \een \ben
                            \hat{t}^{(7)}_{67}
                                  =
960\,{\frac {a+37}{ \left( a+5 \right)  \left( a+4 \right)  \left(
a+3
 \right)  \left( a+2 \right)  \left( a+1 \right) }}
 \een
and formulated our last extrapolation hypothesis.

\begin{conj}
At $j=3$ and at all $N = 3, 4, \ldots$ formula
  \ben
                             \hat{t}^{(N)}_{N-1N}
                                  =
{\frac {(N-3)\,(N-1)!\,(a+7\,N-12)}{3 \left( a+N-2 \right)  \left(
a+N-3 \right) \ldots  \left( a+2 \right) \left( a+1 \right) }}
 \een
defines the last missing off-diagonal  matrix element  of
pseudometric (\ref{kit3}). \label{lemma7}
\end{conj}
\begin{pozn}
We may again display the two subsequent MAPLE-generated quantities,
viz.,
  \ben
                             \hat{t}^{(8)}_{78}
                                  =
8400\,{\frac {a+44}{ \left( a+6 \right)  \left( a+5 \right)  \left(
a+ 4 \right)  \left( a+3 \right)  \left( a+2 \right)  \left( a+1
\right) }}
 \een
and
 \ben \hat{t}^{(9)}_{89}
                                  =
80640\,{\frac {a+51}{ \left( a+7 \right)  \left( a+6 \right)  \left(
a +5 \right)  \left( a+4 \right)  \left( a+3 \right)  \left( a+2
 \right)  \left( a+1 \right) }}\,.
                              \een
The inspection of these formulae demonstrates that our model with
the Laguerre-polynomial wave functions remains
extrapolation-friendly even at $k=3$. This is in contrast with the
observations made in Ref.~\cite{Gegenb} where the complexity of the
elements of metrics grew only too quickly between $k=2$ and  $k=3$.
In {\it loc. cit.}, we even failed to find a reasonable $k=3$
formulae, due to a loss of any obvious guidance for the
extrapolations. In the present model such an overall pattern seems
to stay unchanged. Hence, one could also expect the reasonable
feasibility of further, $k\geq 4$ constructions.
\end{pozn}

\section{Summary and discussion  \label{VII} }

A family of exactly solvable $N-$site quantum lattices with a
non-Hermitian nearest-neighbor interaction was proposed and studied.
The energies appeared real so that each Hamiltonian has been made
self-adjoint, i.e., standard and physical in an {\em ad hoc} Hilbert
space ${\cal H}^{(S)}$ where the inner product was defined via a
metric $\Theta$. The complete set of the eligible metrics has been
shown numbered by a multiindex $(k,\alpha_1,\ldots,\alpha_k)$ in
which the ``degree of non-locality" $k \in \{0,1,\ldots, N-1\}$
indicates that every $\Theta=\Theta_{(k,\vec{\alpha})}$ is a
$(2k+1)-$diagonal matrix. The other free parameters forming the
$k-$plets $\vec{\alpha}$ must only be real and sufficiently small
(otherwise, the metric could cease to be positive definite). In
closed form the metrics were constructed and displayed for
$k=0,k=1,k=2$ and $k=3$.

One of the main merits of the present quantum lattice model lies in
a maximal suppression of its numerical aspects. First of all, the
model does not need numerical methods for the solution of the
time-independent Schr\"{o}dinger equation. The reason is that the
ket-eigenvectors of our special, solvable Hamiltonian $H^{(N)}(a)$
were simply selected in advance (cf. Eqs.~(\ref{jedna}) or
(\ref{eigvec})). We believe that from the point of view of
flexibility of the model such an apparently very strong {\it a
priori} constraint has been more than compensated by the
multiplicity of the eligible metrics
$\Theta_{k,\vec{\alpha}}^{(N)}(a)\neq I$.

In a phenomenological context we explained that our model offers a
new pattern of  a ``smearing" of the position of the lattice sites
in the manner explained in paragraph \ref{physics} above. Such a
form of nonlocality has been shown mediated by the metric. In accord
with the ``weakly nonlocal" interpretation of quantum systems as
advocated in our recent paper \cite{fund} we showed how the ``site"
of the lattice gets smeared over $2k+1$ neighbors, provided only
that the range of smearing $k$ is not too large.

In practice people rarely select {\em both} the suitable Hamiltonian
$H=H^{(input)}$ {\em and} the metric $\Theta=\Theta^{(input)}$ as
the two independent sources of information about quantum dynamics.
An important amendment of this limitation emerged in the literature
cca twelve years ago when a real boom of interest in quantum models
with nontrivial metrics $\Theta \neq I$ has been initiated by Bender
and his colleagues \cite{Carl,BT,BM,BB}. These authors emphasized
that the use of $\Theta \neq I$ seems strongly motivated in quantum
field theory.

Incidentally, in a way explained by Mostafazadeh \cite{cubic} and
emphasized by Jones \cite{Jones} the latter studies were solely
using, in our present language, the long-range metrics
$\Theta^{(N)}_k$ with the maximal possible subscript $k=N-1$. This
made the $k=N-1$ recipe inapplicable in the unitary theory of
scattering \cite{Cannata,Jonesdva}. The core of the problem has been
identified with the locality of the interaction \cite{Alipriv}. The
restoration of the manifest unitarity of quantum scattering has only
been achieved via the use of nonlocal interactions \cite{scatt}.

Fortunately, the difficulties of this type were successfully dealt
with in various chain-interaction phenomenological models (cf.
\cite{D}). A number of constructive $\Theta \neq I$ results appeared
in the context of Bose-Hubbard \cite{Eva} or Friedrichs-Fano-Anderso
\cite{fano}) models. The appeal of these models in condensed-matter
physics resulted in detailed descriptions of the tightly bound
lattices of electrons \cite{Jin}), of the XXZ spin chains
\cite{Korff}) and, last but not least, of certain sophisticated
experiments in optics \cite{Jinbe,Makris}.

\begin{table}[t]
\caption{A sample of the dimension- and parameter-dependence of the
energy spectra $\{E_n(a)\}$ for the quantum-lattice Hamiltonian of
Ref.~\cite{Gegenb}.} \label{pex}

\vspace{2mm}

\centering
\begin{tabular}{||c|c||c|c|c|c|c||}
\hline \hline
 \multicolumn{2}{||c||}{\rm parameters}& \multicolumn{5}{c||}{\rm energies}\\
 \hline
  $N$& $a$ & $E_0(a)$&$E_1(a)$ & \ldots& $E_{N-2}(a)$&$E_{N-1}(a)$\\
 \hline \hline
 6&1.0&
-.6441855418&  -.3709690601& \ldots&
.3709690601& .6441855418
 \\
 &2.0&
-.5622585222&  -.3463446402&\ldots&
 .3463446402&
.5622585222
 \\
 &3.0&
-.5083600312&  -.3244642920&\ldots&
.3244642920& .5083600312
 \\
 \hline
 9&1.0&
-.6443127436&  -.3985980302&\ldots&

 .3985980302&  .6443127436
 \\
 &2.0&
-.5626496595&  -.3794465124& \ldots&
.3794465124&  .5626496595
 \\
 &3.0&
-.5091239690&  -.3616352067& \ldots&
.3616352067&  .5091239690
 \\
 \hline \hline
\end{tabular}
\end{table}

In this context the distinctive feature of our present model lies in
its exact solvability. Another relevant property of our present
model  is that its spectrum is safely real at all $a>0$
(cf.~Eq.~(\ref{secular})). We believe that such a spectrum might
find useful applications, say, in a purely phenomenologically
motivated fitting of measured energies.

Although the similar idea has been also proposed and illustrated,
via another model, in Ref.~\cite{Gegenb}, the present spectrum
(sampled, for reference purposes, in Table~\ref{pexp4} above) looks
much better suited for the fitting purposes. For various values of
parameter $a$ the present energy levels are regularly spread over a
subinterval of the real axis which grows with the growth of the
matrix dimension $N$. In contrast, the distribution of the
bound-state energies of the model of Ref.~\cite{Gegenb} (sampled
here by Table~\ref{pex} for comparison) seems handicapped not only
by its restriction to a fixed interval $(-1,1)$ but, more seriously,
by its not-well-motivated symmetry with respect to its center
(shifted, conveniently, to the origin in Ref.~\cite{Gegenb}) and,
even more seriously, by the weaker sensitivity of its extreme values
to the changes of the dimension parameter $N$.

Besides the above-emphasized potential {\em physical} relevance of
the exact solvability of our present family of Laguerrian
bound-state models we would like to mention, in the conclusion, also
the purely mathematical appeal and consequences of our closed-form
constructions.

Firstly, the unusual though still fully non-numerical solvability of
our model could serve, in principle at least, in the role of the
initial zeroth approximation in perturbation theory. After all, just
a very few exactly solvable models with metrics $\Theta \neq I$
exist on the market. More attention has only been paid to a few
exceptionally simple quantum lattices with certain extremely
elementary point-like interactions \cite{metriky,...}.

Secondly, one should emphasize that it was certainly unexpected that
the innocent-looking freedom in the choice of the proportionality
constants $\kappa_n^{(N)}(a)$ in relation (\ref{kappas}) proved able
to render the general definition of metric (\ref{hu}) compatible
with the fairly strong {\em simplification} requirement of the
sparse-matrix structure of its  $(2k+1)-$diagonal representations
$\Theta_{k,\vec{\alpha}}^{(N)}(a)$ as specified by Eqs.~(\ref{kit})
and (\ref{every}).

From the point of view of the theory of classical orthogonal
polynomials the latter result implies that every set of column
vectors (\ref{eigvec}) of Laguerre polynomials may be assigned {\em
many different} biorthogonalized sets of row vectors. They may be
formed of the ``ketket" eigenvectors $|\psi\kkt$ of $\left
[H^{(N)}(a)\right ]^\dagger$ and classified by the same multiindex
$\{k,\vec{\alpha}\}$ as the metrics, therefore. Indeed, we have
 \ben
 |\psi_n^{(N)}(a)\kkt=
 \Theta_{k,\vec{\alpha}}^{(N)}(a)\,|\psi_n^{(N)}(a)\kt
 \een
or, in the componentwise notation of section \ref{physics},
  \be
  \left (|\psi_n^{(N)}(a)\kkt\right )_s=
  \sum_{s'={\rm max}(1,s-k)}^{{\rm min}(N,s+k)}\,
  \left (\Theta_{k,\vec{\alpha}}^{(N)}(a)
   \right )_{ss'}
   \left (|\psi_n^{(N)}(a)\kt
  \right )_{s'}
  \,
  \ee
where $s=1,2,\ldots,N$ and where matrices
$\Theta_{k,\vec{\alpha}}^{(N)}(a)$ are now at our disposal in closed
form (\ref{every}), for $0 \leq k \leq 3$ at least.

Thirdly, the variability of the $k-$parametric $(2k+1)-$diagonal
metrics $\Theta_{k,\vec{\alpha}}^{(N)}(a)$ could prove useful during
the model-building in which the  energies are prescribed by our
``Laguerrean" input Hamiltonian $H^{(N)}(a)$ but in which {\em
several other} matrices of observables might be required selfadjoint
in at least one of the eligible Hilbert spaces of states ${\cal
H}_{k,\vec{\alpha}}^{(S)}$, typically, via the fitting of the $k$
free parameters $\vec{\alpha}$.

Fourthly, on the level of methods the present approach to
solvability could open new ways of circumventing the difficulties
which emerged during the study of scattering with $H\neq H^\dagger$
\cite{Jones,Cannata}. The use of the non-numerical, exactly solvable
models with tridiagonal Hamiltonians and large $N \to \infty$ seems
to have been quite effective in this context \cite{D,SIGMAscatt}.

Fifthly, the  quick progress achieved during the study of  ${\cal
PT-}$symmetric solvable differential equations~\cite{Levai} has not
been followed by the sufficiently rapid progress in understanding of
the related metrics. One of the reasons lies in the virtually
prohibitive technical obstacles \cite{cubic}. In contrast,  the
recent transition to  ${\cal PT-}$symmetric difference
Schr\"{o}dinger equations has been accompanied by the comparatively
quick success in finding the comparatively extensive sets of metrics
$ \Theta$ compatible with a given $H$
\cite{fund,SIGMAscatt,fundgra}.

Sixthly, a deeper understanding of the problem of ambiguity of the
assignment of a metric to an input Hamiltonian has been reached here
via the matrix Schr\"{o}dinger equations. We pointed out that the
old paradoxes \cite{Geyer} are finding new resolutions, say, due to
the feasibility of a ``smearing" of the coordinates at $N < \infty$.
Due to the variability of $k$ in our model we showed how the smeared
localization could be studied by the experimental measurements, in
principle at least (cf. also the possible application of this idea
to cryptohermitian guantum graphs as mentioned in
Refs.~\cite{fundgra,anomal}).

Last but not least, our continuing attention paid to the exactly
solvable models could clarify the possibilities of the practical use
of the concept of the hidden Hermiticity (cryptohermiticity) also in
connection with the unitary time evolution where the time-dependent
metrics emerge. As long as this idea leads to many new technical
challenges \cite{timedep}, our present non-numerical model could
re-demonstrate, in the nearest future, its relevance and importance
also in this new context.


\section*{Acknowledgement}

Work supported by the GA\v{C}R grant Nr. P203/11/1433, by the
M\v{S}MT ``Doppler Institute" project Nr. LC06002 and by the
Institutional Research Plan AV0Z10480505.

\newpage

\section*{Appendix: Three Hilbert spaces ${\cal H}^{(F)}$,
${\cal H}^{(S)}$ and  ${\cal H}^{(P)}$ and the concept of hidden
Hermiticity}

Whenever a non-Hermitian Hamiltonian matrix $H \neq H^\dagger$ with
real spectrum is declared  physical, the Hilbert space of states
{\em cannot} be represented by the most common $N-$dimensional
version of $\ell^2$. The latter space may only play an auxiliary
role (this space without any immediate physical interpretation will
be denoted here by the symbol ${\cal H}^{(F)}$ where the superscript
$^{(F)}$ may stand for ``friendly" as well as for ``false"
\cite{SIGMA}).

For the work with Hamiltonians  $H \neq H^\dagger$ there exists a
trick introduced in physics, presumably, by Scholtz et al
\cite{Geyer}. Its essence lies in the replacement of  inappropriate
${\cal H}^{(F)}$ by a unitarily non-equivalent, ``standardized"
Hilbert space ${\cal H}^{(S)}$. By construction, the two Hilbert
spaces share the same set of kets (in our notation, $|\psi\kt \in
{\cal V}^{(S)}={\cal V}^{(F)}={\cal V}$). The only distinguishing
feature is that the usual Hermitian conjugation, i.e., the most
conventional transition to the duals, ${\cal T}^{(F)}:\,{\cal V}\to
{\cal V}'$ (mapping
 \be
 {\cal T}^{(F)}: \ |\psi \kt \ \to \
   \br \psi|\,\ \ \ \ \ \ {\rm in }\ \ \ \ \ {\cal H}^{(F)}
   \label{dirovo}
 \ee
in the current Dirac's notation) must be replaced, in ${\cal
H}^{(S)}$, by the generalized conjugation ${\cal T}^{(S)}: \,{\cal
V}\,(={\cal V}^{(S)}\,)\to \left [{\cal V}'\right ]^{(S)}$. The
latter mapping generates the different duals defined in terms of a
suitable matrix $\Theta$ of metric~\cite{SIGMA},
 \be
 {\cal T}^{(S)}: \ |\psi \kt \ \to \
   \br \psi^{(S)}|:= \bbr \psi|:=\br \psi|\,\Theta
   \,\ \ \ \ \ \ {\rm in }\ \ \ \ \ {\cal H}^{(S)}\,.
   \label{newcon}
 \ee
An instructive illustration of this scenario may be found in paper
\cite{Geyer} where the kets $|\psi\kt\in {\cal V}={\cal
V}^{(F)}={\cal V}^{(S)}$ were just bosonic, imperfect
representatives of the true nuclear (i.e., {\em fermionic}) state
vectors
 \be
 |\psi\pkt = \Omega\,|\psi\kt\ \in \ {\cal W}\,
 \label{newd}
 \ee
which lied in the textbook Hilbert space ${\cal H}^{(P)}$ with
metric $\Theta^{(P)}=I$. The so called Dyson's map $\Omega: {\cal V}
\to {\cal W}$ of bosons upon fermions (\ref{newd}) was chosen
non-unitary so that $\pbr \psi_a|\psi_b\pkt \neq  \br
\psi_a|\psi_b\kt$ since $\Omega^\dagger \neq \Omega^{-1}$.

In opposite direction the non-unitarity of the map $\Omega$ enables
us to require that
 \be
 \pbr \psi_a|\psi_b\pkt =
 \br \psi_a|\Omega^\dagger \Omega|\psi_b\kt=
 \bbr \psi_a|\psi_b\kt\,.
 \ee
This formula connects the Dyson's map and the metric by the most
important relation
 \be
 \Theta=\Omega^\dagger \Omega\,.
 \label{factorized}
 \ee
The same picture of reality is obtained in {\em both} of the
alternative physical Hilbert spaces of states ${\cal H}^{(P)}$ and
${\cal H}^{(S)}$ which are, by construction, unitarily equivalent.

In the light of definition (\ref{newd}) the result of action of $H$
upon $|a\kt$, i.e., a new vector $|b\kt=H|a\kt$ appears in
equivalent relation $\Omega^{-1}|b\pkt=H\Omega^{-1}|a\pkt,$ i.e., we
have $|b\pkt=\mathfrak{h}|a\pkt$ where we abbreviated
$\mathfrak{h}=\Omega H\Omega^{-1}$. The latter image of Hamiltonian
must be self-adjoint in ${\cal H}^{(P)}$, i.e., the operator
$\mathfrak{h}=\Omega H\Omega^{-1}$ must be equal to its conjugate in
${\cal H}^{(P)}$, viz., to $\mathfrak{h}^\dagger=\left
(\Omega^{-1}\right )^\dagger H^\dagger\Omega^\dagger$. Such a
constraint imposed in ${\cal H}^{(P)}$ is strictly equivalent to
formula (\ref{dieu}) valid in the other two spaces ${\cal H}^{(S)}$
and ${\cal H}^{(F)}$.

For the Hamiltonian operator we have to distinguish between its
apparent non-Hermiticity $H\neq H^\dagger$ in unphysical space
${\cal H}^{(F)}$ and the true, ``hidden" Hermiticity in the physical
Hilbert space ${\cal H}^{(S)}$ a.k.a. cryptohermiticity.  In the
latter space the Hermitian conjugate of $H$ (let us denote it by the
symbol $H^\ddagger$) is defined, consistently, by the prescription
 \be
 H^\ddagger:=\Theta^{-1}H^\dagger\,\Theta\,\ \ \ \ {\rm in}\ \ \
 {\cal H}^{(S)}\,.
 \label{anomalous}
 \ee
In this language, Eq.~(\ref{dieu}) is precisely the disguised
condition $H = H^\ddagger$ of the Hermiticity of the Hamiltonian at
a fixed $\Theta$. This scenario has been used, e.g., in
Ref.~\cite{Geyer}. On the contrary, whenever we select the concrete
form of matrix $H$ in advance, relations~(\ref{dieu}) must be reread
as the Dieudonn\'{e}'s (incomplete) set of linear equations for the
matrix elements of $\Theta=\Theta(H)$. This is precisely the
approach used in our present paper.

\end{document}